\documentclass[letterpaper,aps,prl,superscriptaddress,showpacs,floatfix,twocolumn]{revtex4}

\usepackage{graphicx}	% Include figure filed
\usepackage{xspace}	% Include xspace

\newcommand{\pt}{\mbox{$p_T$}\xspace}

\begin{document}

\title{Cross Section and Parity Violating Spin Asymmetries of $W^{\pm}$ Boson Production
 in Polarized $p+p$ Collisions at $\sqrt{s}=500$ GeV}

\newcommand{\abilene}{Abilene Christian University, Abilene, Texas 79699, USA}
\newcommand{\banaras}{Department of Physics, Banaras Hindu University, Varanasi 221005, India}
\newcommand{\barc}{Bhabha Atomic Research Centre, Bombay 400 085, India}
\newcommand{\bnlcoll}{Collider-Accelerator Department, Brookhaven National Laboratory, Upton, New York 11973-5000, USA}
\newcommand{\bnlphys}{Physics Department, Brookhaven National Laboratory, Upton, New York 11973-5000, USA}
\newcommand{\caucr}{University of California - Riverside, Riverside, California 92521, USA}
\newcommand{\charlesczech}{Charles University, Ovocn\'{y} trh 5, Praha 1, 116 36, Prague, Czech Republic}
\newcommand{\chonbuk}{Chonbuk National University, Jeonju, 561-756, Korea}
\newcommand{\ciae}{China Institute of Atomic Energy (CIAE), Beijing, People's Republic of China}
\newcommand{\cns}{Center for Nuclear Study, Graduate School of Science, University of Tokyo, 7-3-1 Hongo, Bunkyo, Tokyo 113-0033, Japan}
\newcommand{\colorado}{University of Colorado, Boulder, Colorado 80309, USA}
\newcommand{\columbia}{Columbia University, New York, New York 10027 and Nevis Laboratories, Irvington, New York 10533, USA}
\newcommand{\czechtech}{Czech Technical University, Zikova 4, 166 36 Prague 6, Czech Republic}
\newcommand{\dapnia}{Dapnia, CEA Saclay, F-91191, Gif-sur-Yvette, France}
\newcommand{\elte}{ELTE, E{\"o}tv{\"o}s Lor{\'a}nd University, H - 1117 Budapest, P{\'a}zm{\'a}ny P. s. 1/A, Hungary}
\newcommand{\ewha}{Ewha Womans University, Seoul 120-750, Korea}
\newcommand{\fsu}{Florida State University, Tallahassee, Florida 32306, USA}
\newcommand{\gsu}{Georgia State University, Atlanta, Georgia 30303, USA}
\newcommand{\hanyang}{Hanyang University, Seoul 133-792, Korea}
\newcommand{\hiroshima}{Hiroshima University, Kagamiyama, Higashi-Hiroshima 739-8526, Japan}
\newcommand{\ihepprot}{IHEP Protvino, State Research Center of Russian Federation, Institute for High Energy Physics, Protvino, 142281, Russia}
\newcommand{\illuiuc}{University of Illinois at Urbana-Champaign, Urbana, Illinois 61801, USA}
\newcommand{\instpasczech}{Institute of Physics, Academy of Sciences of the Czech Republic, Na Slovance 2, 182 21 Prague 8, Czech Republic}
\newcommand{\isu}{Iowa State University, Ames, Iowa 50011, USA}
\newcommand{\jinrdubna}{Joint Institute for Nuclear Research, 141980 Dubna, Moscow Region, Russia}
\newcommand{\jyvaskyla}{Helsinki Institute of Physics and University of Jyv{\"a}skyl{\"a}, P.O.Box 35, FI-40014 Jyv{\"a}skyl{\"a}, Finland}
\newcommand{\kek}{KEK, High Energy Accelerator Research Organization, Tsukuba, Ibaraki 305-0801, Japan}
\newcommand{\kfki}{KFKI Research Institute for Particle and Nuclear Physics of the Hungarian Academy of Sciences (MTA KFKI RMKI), H-1525 Budapest 114, POBox 49, Budapest, Hungary}
\newcommand{\korea}{Korea University, Seoul, 136-701, Korea}
\newcommand{\kurchatov}{Russian Research Center ``Kurchatov Institute", Moscow, Russia}
\newcommand{\kyoto}{Kyoto University, Kyoto 606-8502, Japan}
\newcommand{\labllr}{Laboratoire Leprince-Ringuet, Ecole Polytechnique, CNRS-IN2P3, Route de Saclay, F-91128, Palaiseau, France}
\newcommand{\lawllnl}{Lawrence Livermore National Laboratory, Livermore, California 94550, USA}
\newcommand{\losalamos}{Los Alamos National Laboratory, Los Alamos, New Mexico 87545, USA}
\newcommand{\lpc}{LPC, Universit{\'e} Blaise Pascal, CNRS-IN2P3, Clermont-Fd, 63177 Aubiere Cedex, France}
\newcommand{\lund}{Department of Physics, Lund University, Box 118, SE-221 00 Lund, Sweden}
\newcommand{\maryland}{University of Maryland, College Park, Maryland 20742, USA}
\newcommand{\mass}{Department of Physics, University of Massachusetts, Amherst, Massachusetts 01003-9337, USA }
\newcommand{\muenster}{Institut fur Kernphysik, University of Muenster, D-48149 Muenster, Germany}
\newcommand{\muhlenberg}{Muhlenberg College, Allentown, Pennsylvania 18104-5586, USA}
\newcommand{\myongji}{Myongji University, Yongin, Kyonggido 449-728, Korea}
\newcommand{\nagasaki}{Nagasaki Institute of Applied Science, Nagasaki-shi, Nagasaki 851-0193, Japan}
\newcommand{\newmex}{University of New Mexico, Albuquerque, New Mexico 87131, USA }
\newcommand{\nmsu}{New Mexico State University, Las Cruces, New Mexico 88003, USA}
\newcommand{\ohio}{Department of Physics and Astronomy, Ohio University, Athens, OH 45701, USA}
\newcommand{\ornl}{Oak Ridge National Laboratory, Oak Ridge, Tennessee 37831, USA}
\newcommand{\orsay}{IPN-Orsay, Universite Paris Sud, CNRS-IN2P3, BP1, F-91406, Orsay, France}
\newcommand{\peking}{Peking University, Beijing, People's Republic of China}
\newcommand{\pnpi}{PNPI, Petersburg Nuclear Physics Institute, Gatchina, Leningrad region, 188300, Russia}
\newcommand{\riken}{RIKEN Nishina Center for Accelerator-Based Science, Wako, Saitama 351-0198, JAPAN}
\newcommand{\rikjrbrc}{RIKEN BNL Research Center, Brookhaven National Laboratory, Upton, New York 11973-5000, USA}
\newcommand{\rikkyo}{Physics Department, Rikkyo University, 3-34-1 Nishi-Ikebukuro, Toshima, Tokyo 171-8501, Japan}
\newcommand{\saispbstu}{Saint Petersburg State Polytechnic University, St. Petersburg, Russia}
\newcommand{\saopaulo}{Universidade de S{\~a}o Paulo, Instituto de F\'{\i}sica, Caixa Postal 66318, S{\~a}o Paulo CEP05315-970, Brazil}
\newcommand{\seoulnat}{Seoul National University, Seoul, Korea}
\newcommand{\stonybrkc}{Chemistry Department, Stony Brook University, SUNY, Stony Brook, New York 11794-3400, USA}
\newcommand{\stonycrkp}{Department of Physics and Astronomy, Stony Brook University, SUNY, Stony Brook, New York 11794-3400, USA}
\newcommand{\tenn}{University of Tennessee, Knoxville, Tennessee 37996, USA}
\newcommand{\titech}{Department of Physics, Tokyo Institute of Technology, Oh-okayama, Meguro, Tokyo 152-8551, Japan}
\newcommand{\tsukuba}{Institute of Physics, University of Tsukuba, Tsukuba, Ibaraki 305, Japan}
\newcommand{\vandy}{Vanderbilt University, Nashville, Tennessee 37235, USA}
\newcommand{\weizmann}{Weizmann Institute, Rehovot 76100, Israel}
\newcommand{\yonsei}{Yonsei University, IPAP, Seoul 120-749, Korea}
\affiliation{\abilene}
\affiliation{\banaras}
\affiliation{\barc}
\affiliation{\bnlcoll}
\affiliation{\bnlphys}
\affiliation{\caucr}
\affiliation{\charlesczech}
\affiliation{\chonbuk}
\affiliation{\ciae}
\affiliation{\cns}
\affiliation{\colorado}
\affiliation{\columbia}
\affiliation{\czechtech}
\affiliation{\dapnia}
\affiliation{\elte}
\affiliation{\ewha}
\affiliation{\fsu}
\affiliation{\gsu}
\affiliation{\hanyang}
\affiliation{\hiroshima}
\affiliation{\ihepprot}
\affiliation{\illuiuc}
\affiliation{\instpasczech}
\affiliation{\isu}
\affiliation{\jinrdubna}
\affiliation{\jyvaskyla}
\affiliation{\kek}
\affiliation{\kfki}
\affiliation{\korea}
\affiliation{\kurchatov}
\affiliation{\kyoto}
\affiliation{\labllr}
\affiliation{\lawllnl}
\affiliation{\losalamos}
\affiliation{\lpc}
\affiliation{\lund}
\affiliation{\maryland}
\affiliation{\mass}
\affiliation{\muenster}
\affiliation{\muhlenberg}
\affiliation{\myongji}
\affiliation{\nagasaki}
\affiliation{\newmex}
\affiliation{\nmsu}
\affiliation{\ohio}
\affiliation{\ornl}
\affiliation{\orsay}
\affiliation{\peking}
\affiliation{\pnpi}
\affiliation{\riken}
\affiliation{\rikjrbrc}
\affiliation{\rikkyo}
\affiliation{\saispbstu}
\affiliation{\saopaulo}
\affiliation{\seoulnat}
\affiliation{\stonybrkc}
\affiliation{\stonycrkp}
\affiliation{\tenn}
\affiliation{\titech}
\affiliation{\tsukuba}
\affiliation{\vandy}
\affiliation{\weizmann}
\affiliation{\yonsei}
\author{A.~Adare} \affiliation{\colorado}
\author{S.~Afanasiev} \affiliation{\jinrdubna}
\author{C.~Aidala} \affiliation{\losalamos}
\author{N.N.~Ajitanand} \affiliation{\stonybrkc}
\author{Y.~Akiba} \affiliation{\riken} \affiliation{\rikjrbrc}
\author{R.~Akimoto} \affiliation{\cns}
\author{J.~Alexander} \affiliation{\stonybrkc}
\author{H.~Al-Taani} \affiliation{\nmsu}
\author{K.R.~Andrews} \affiliation{\abilene}
\author{A.~Angerami} \affiliation{\columbia}
% ...
\collaboration{PHENIX Collaboration} \noaffiliation

\date{\today}

\begin{abstract}
Large parity violating longitudinal single-spin asymmetries 
$A^{e^+}_L= -0.86^{+0.30}_{-0.14}$ 
and $A^{e^-}_L= 0.88^{+0.12}_{-0.71}$
are observed for inclusive high transverse momentum
electrons and positrons
in polarized $p+p$ collisions at a center of mass energy
of $\sqrt{s}=500\!$ GeV with the PHENIX detector at RHIC.
These $e^{\pm}$ come mainly from the decay of $W^{\pm}$ and 
$Z^0$ bosons, and their asymmetries 
directly demonstrate parity violation in the couplings of the 
$W^{\pm}$ to the light quarks.
The observed electron and positron yields were used to estimate
$W^\pm$ boson production cross sections 
for the $e^\pm$ channels of
$\sigma(pp\!\rightarrow\!W^{+}X)\times BR(W^+\!\rightarrow\!e^+\nu_e)=144.1\pm 
21.2({\rm stat})^{+3.4}_{-10.3}({\rm syst})\pm15\%(norm)$~pb,
and $\sigma(pp\!\rightarrow\!W^{-}X)\times 
BR(W^-\!\rightarrow\!e^-\bar{\nu_e})=31.7\pm 
12.1({\rm stat})^{+10.1}_{-8.2}({\rm syst})\pm15\%(norm)$~pb.

\end{abstract}

% It is optional to also add (uncomment):
% \keywords{}

\maketitle
% Begin paper body
 Determining the contributions of the partons to the spin of the proton
 is a crucial element in our understanding of QCD~\cite{Jaffe:1989jz,Leader:1988vd,Kuhn:2008sy}.
 Polarized inclusive deep inelastic scattering
experiments (DIS) have measured the combination of 
valence and sea quark, and gluon helicity distributions 
($\Delta q+\Delta\bar{q}$, $\Delta g$)~\cite{deFlorian:2009vb,Kuhn:2008sy}.
Analysis of polarized semi-inclusive DIS experiments~\cite{Alekseev:2010,Airapetian:2004zf,Adeva:1997qz}
have determined the individual flavor separated $\Delta q$ and 
$\Delta \bar{q}$ by connecting final state hadrons with quark 
flavors using fragmentation functions. 
Collisions of longitudinally polarized protons at high energies 
allow study of $\Delta g$~\cite{deFlorian:2009vb,Adare:2008px,Abelev:2007vt}, and can provide 
complementary measurements of $\Delta u$, $\Delta \overline{u}$, 
$\Delta d$, $\Delta \overline{d}$~\cite{Bunce:2000uv,Bourrely:1993dd}. 
In particular, the production of $W^{\pm}$ 
couples only the left-handed quarks and right-handed antiquarks 
($u_L \bar{d}_{R} \rightarrow W^{+}$ and $d_L \bar{u}_{R} \rightarrow W^{-}$), 
so the asymmetry of the $W$ yield from flipping the helicity of a 
polarized proton is sensitive to the flavor dependence of $\Delta q$ 
and $\Delta \bar{q}$. 
Production of 
the $W$ occurs at a scale where higher order QCD corrections 
can be evaluated reliably, and it is free from uncertainties 
in fragmentation functions by detecting leptons from $W$ 
decay~\cite{deFlorian:2010aa,Nadolsky:2003ga}.

The first observations of $W$-boson production in 
polarized $p+p$ collisions, and direct demonstration of the 
parity-violating coupling of the $W$ to the light quarks 
are reported here by PHENIX and in a companion paper by 
STAR~\cite{STAR_W} for $\sqrt{s}=500$ GeV at the Relativistic
Heavy Ion Collider.
The 2009 PHENIX data are from polarized
$pp \rightarrow e^{\pm} + X$, where the $e^{\pm}$ with $p_{T}\!>\!30$~GeV/$c$ 
come mainly from $W$ and $Z$ decays. 

The PHENIX detector has been described in detail elsewhere~\cite{Adcox:2003zm}.
This analysis is based on data collected with two central arm spectrometers, 
each covering $|\Delta\phi|<\pi/2$ in azimuth and
$\left|\eta\right|<0.35$ in pseudorapidity, which surround the central
axial magnetic field.  
The bend angle of charged tracks is determined by drift chambers outside the magnetic field 
starting at a radius of 2.02 m from the beamline.
The longitudinal position $z$ of the track is measured as it exits 
the drift chamber in a layer of pad chambers 2.46 m from the beamline
with a spatial resolution of $\sigma_{z} = 1.7$ mm.
The electromagnetic calorimeter, located at a radial distance of 
$\sim$5 m from the beam line, 
is used to measure the energy, position, and time of flight of electrons.
In this analysis,
the $\pt$ dependence of the reconstructed $\pi^{0}$ and $\eta$ mass peaks
was used to confirm the energy scale and linearity to within 2.5\%.  
The $p_T$ dependence of the peak widths was used to determine the energy resolution 
$\sigma_{E}/E = 8.1\%/\sqrt{E} (GeV) \bigoplus 5.0\%$.

A trigger with a nominal 10 GeV threshold 
in the electromagnetic calorimeter selected events for this analysis.
This trigger was fully efficient for $e^{\pm}$ with 
transverse momentum $\pt$ above 12~GeV/$c$.
Charged tracks reconstructed in the drift
chambers and the pad chambers which match the calorimeter
cluster with $\left|\Delta\phi\right|<0.01$ were used to reconstruct
the $z$ position of the event vertex. Only events 
with $\left|z\right|<30$ cm that are well within 
the acceptance of the central arm spectrometers were analyzed.
Loose cuts on the time of flight measured by the calorimeter and energy-momentum matching
suppressed accidental matches and cosmic rays.

The analyzed data sample corresponds to 
an integrated luminosity of 8.6 pb$^{-1}$, which was
determined from beam-beam counter coincidences and corrected for
a small (6\%) effect from multiple collisions per beam crossing. 
The beam-beam counters are two 64 channel quartz \v{C}erenkov counters $\pm$1.44 m from the 
center of the detector and cover a pseudorapidity range of 3.1$<\left|\eta\right|<$3.9.
The cross section for coincidences within $|z|\lesssim 30$ cm was found to be $32.5 \pm 3.2$ mb from
the van der Meer scan technique~\cite{Adare:2008qb}.

%%%%%%%%%%%%%%%%%%%%%%%%%%%%%%%%%%%%%%%%%%%%%%%%%%%%%%%%%%%%%%%  Fig_1
\begin{figure}[thb]
\includegraphics[width=1.05\linewidth]{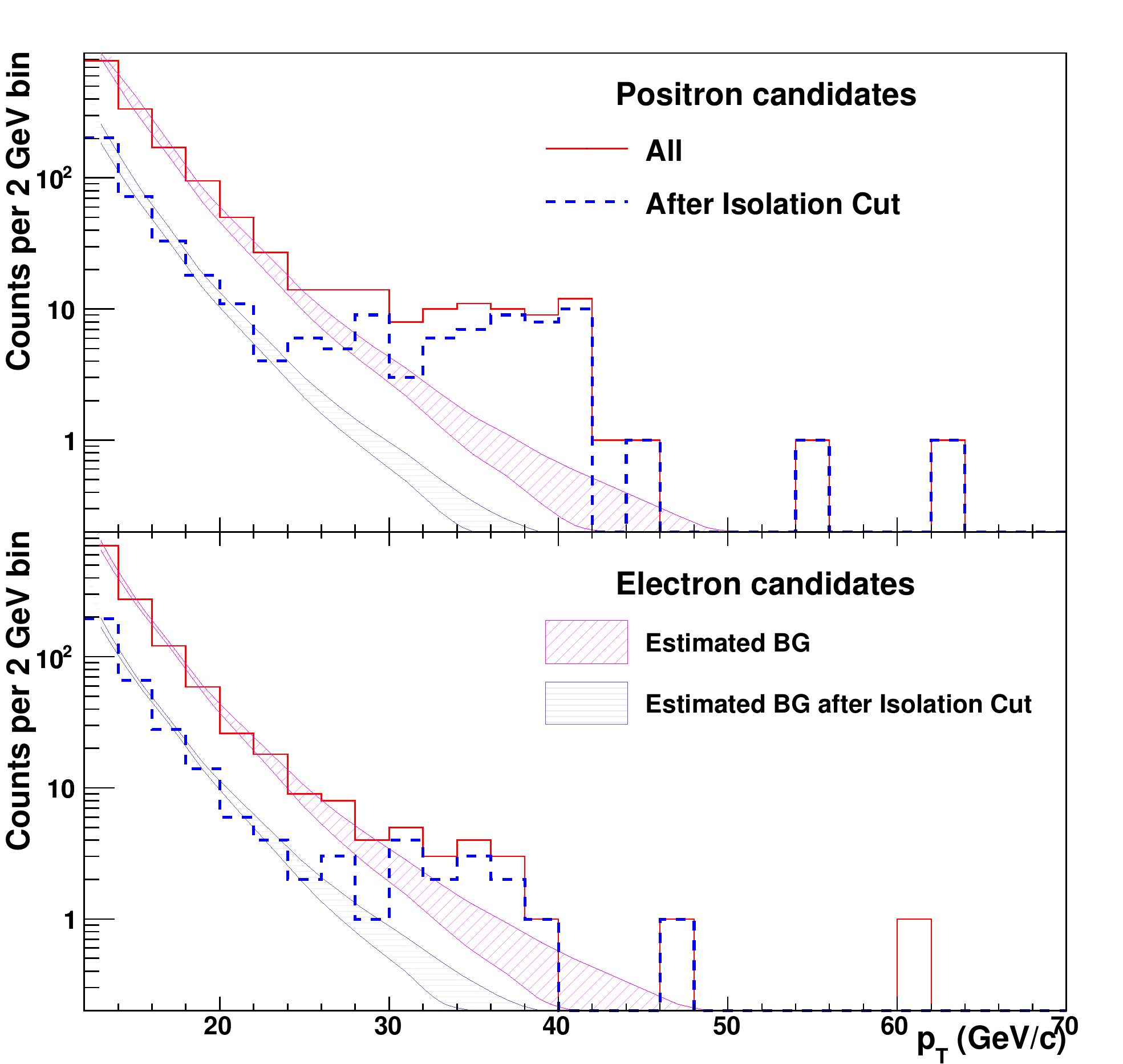}
\caption{(color online) The spectra of positive (upper panel) and negative (lower panel) candidates 
before (solid histogram) and after (dashed histogram) an isolation cut. 
The estimated background bands are also shown.
The computation of the background before the isolation cut is described in the text.
The background band after the isolation cut is computed by scaling the background 
before the isolation cut by the isolation cut efficiency measured in 
the background region ($12<p_{T}<20$ GeV/$c$). 
}
\label{fig:cy_p}
\end{figure}

The resulting yield of positive and negative electron candidates is shown
in Fig.~\ref{fig:cy_p} 
where $\pt$ has been determined from the calorimeter cluster energy.
The charge sign is determined from the bend angle, $\alpha$, measured in the drift 
chamber, and the nominal transverse beam position.
The angular resolution and stability of beam position were monitored by frequent runs with no magnetic field.
The resolution $\sigma_{\alpha}$ was typically about 1.1 mr, to be compared to 
a 2.3 mr bend angle for 40~GeV/$c$ tracks.
The variation in the average transverse beam position measured by reconstruction of the primary vertex
in these runs was within $\pm300\,\mu$m, 
and did not affect the charge determination.
The probability of charge misidentification at 40 GeV/$c$ was estimated to be
less than 2\%.

In addition to $e^{\pm}$ from $W$ and $Z$ decay, this sample of events contains
various backgrounds.
The dominant backgrounds were photon conversions before the drift chamber 
and charged hadrons.
These were estimated using the raw calorimeter cluster distribution and the charged pion spectra predicted
by perturbative QCD
convoluted with the hadronic response of the calorimeter tuned to reproduce test beam data. 
This calculated background was normalized to the measured spectrum in the region $12\!<\!\pt\!<\!20$ GeV/$c$
and extrapolated to higher $\pt$.
Electrons from heavy flavor decay were estimated from an FONLL calculation~\cite{Cacciari:2005rk} which agrees
well with the prompt electron measurement at  $\sqrt{s}=200$ GeV\cite{Adare:2006hc}. PYTHIA~\cite{Sjostrand:2000wi} 
was used to estimate the contributions of electrons with $p_T\!>\!30$ GeV/$c$ 
from $\tau$ lepton decays of $W$ and $Z$ bosons. These two components were found to be negligible.
The background bands in Fig.~\ref{fig:cy_p}  
include uncertainties in the photon 
conversion probability, the background normalization, 
and the background extrapolation to $p_{T}\!>\!30$ GeV/$c$. 

The acceptance was calculated from Monte Carlo simulation and corrected for time dependent variations in tracking and
trigger efficiency measured with a minimum bias data sample triggered by beam-beam counter coincidences.
The reconstruction efficiency for tracks within the geometrical
acceptance of the spectrometer
is the product of the active areas of the drift chamber and 
the pad chamber (54\%), a fiducial cut on the calorimeter 
to ensure the showers are not near an edge (81\%),
and a fiducial cut on the drift chamber to avoid charge misidentification (85\%).
The reconstruction efficiency was not $\pt$ dependent for $\pt >$30 GeV/$c$.

%%%%%%%%%%%%%%%%%%%%%%%%%%%%%%%%%%%%%%%%%%%%%%%%%%%%%%%%%%%%%%%  Fig_2
\begin{figure}[t]
\includegraphics[width=1.0\linewidth]{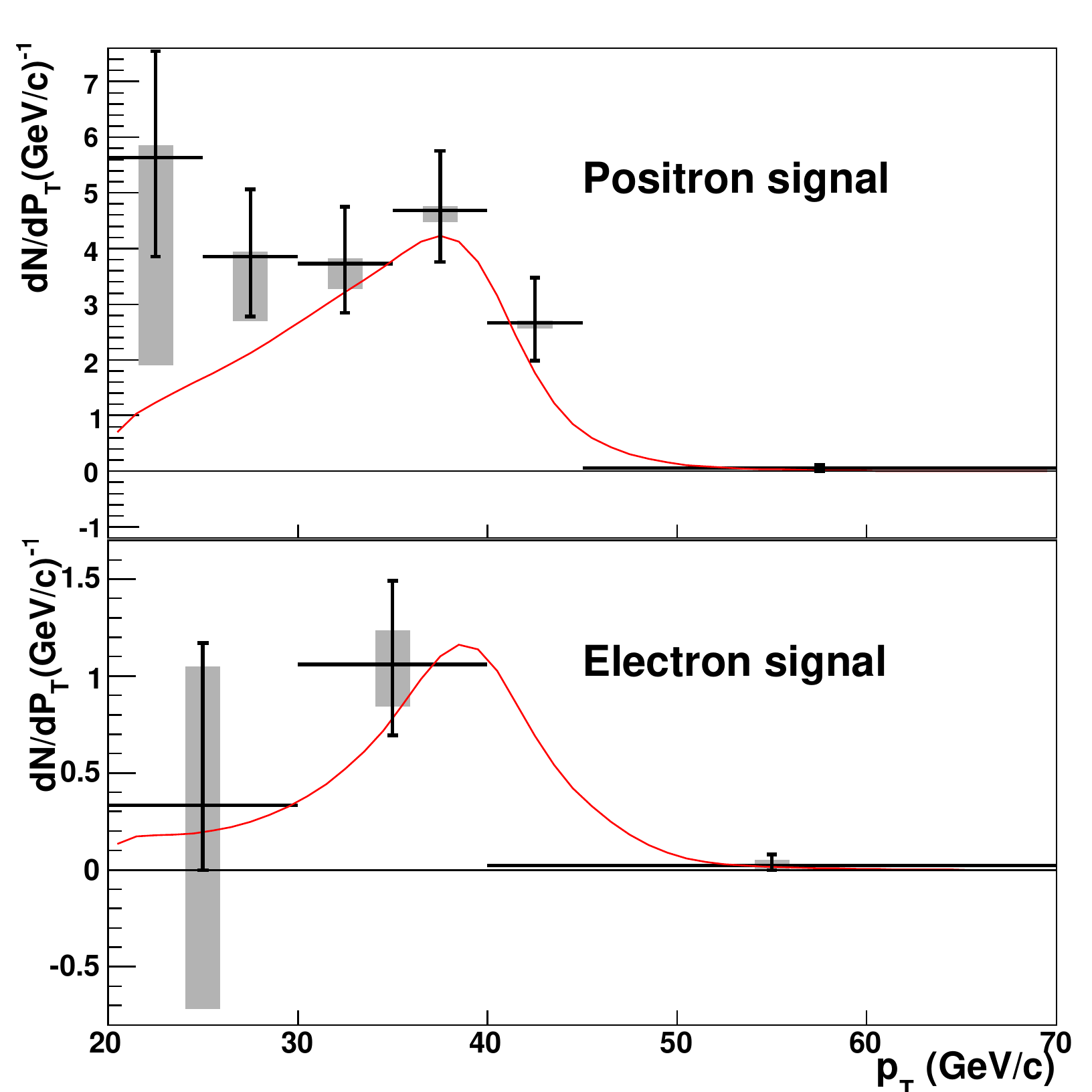}
\caption{(color online) 
Background subtracted spectra of positron (upper panel) and electron 
(lower panel) candidates taken from all counts compared to the spectrum of 
$W$ and $Z$ decays from an NLO 
calculation~\cite{deFlorian:2010aa,Nadolsky:2003ga} (see text).
The gray bands reflect the range of background estimates.}
\label{fig:cs_p}
\end{figure}

Figure \ref{fig:cs_p} shows the background subtracted signal
in the acceptance of the detector
for positive and negative charges respectively compared to the 
spectrum predicted by NLO calculations~\cite{deFlorian:2010aa,Nadolsky:2003ga,Melnikov:2006kv} 
normalized for the integrated luminosity, corrected for the detector efficiency and acceptance, 
and smeared by the energy resolution of the calorimeter.
The yield measured by counting events in the signal ($30\!<\!\pt\!<\!50$ GeV/$c$) region 
is compared to the predicted yield in Table \ref{table:xsecresult}.
The measured yield in the signal region
is consistent with the predictions of the NLO 
and NNLO~\cite{Melnikov:2006kv} calculations.

\begin{figure}[t]
\includegraphics[width=1.0\linewidth]{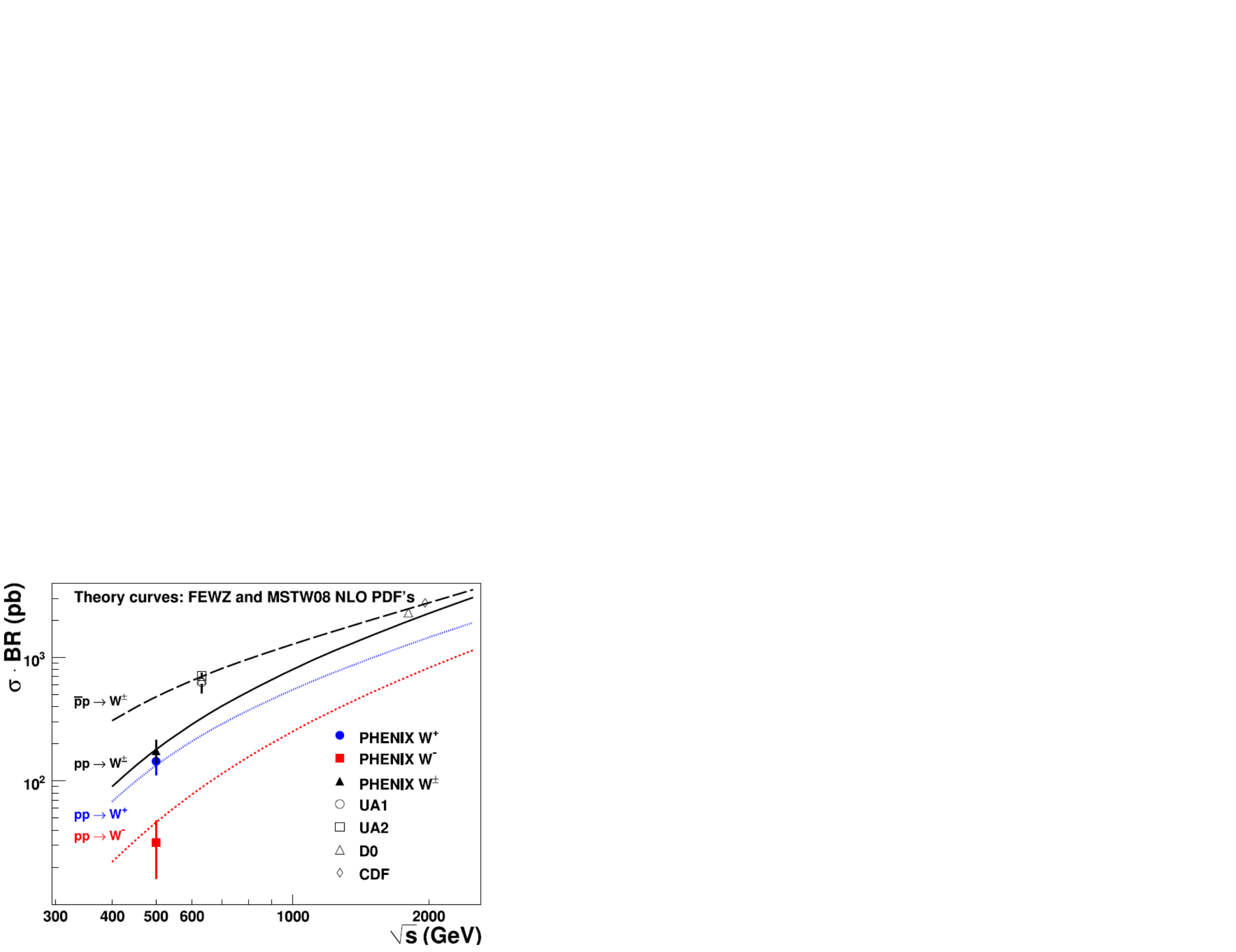}
\caption{(color online) 
Inclusive cross sections for $W$ leptonic decay channel 
of this measurement and $\bar{p}p$ measurements 
~\cite{Acosta:2004uq,Abbott:1999tt,Alitti:1990gj,Albajar:1988ka}. 
Statistical and systematic uncertainties were added here in quadrature.
Curves represent theory calculations.
}
\label{fig:xsecfig}
\end{figure}

To extract the $W^{\pm}$ production cross section, we estimated the $Z$
fraction in our sample with the NLO and NNLO calculations and
MRST~\cite{Martin:2002aw} and MSTW~\cite{Martin:2009iq} PDFs.
The contribution from $Z$ decays is approximately $\sim$7\% for $W^+$ and $\sim$30\% 
for $W^-$. The fraction of the total cross section within 
$|y|\!<\!0.35$, $p_{T}\!>\!30$ GeV/$c$, and $|\Delta \phi|\!<\!\pi$
is estimated to be $\sim$11\% of positrons from $W^+$
and $\sim$7.5\% of electrons from $W^-$ from these calculations.
The variation of the calculation is small compared to other 
sources of systematic uncertainty.
With these corrections, 
$\sigma(pp\rightarrow W^{+}X)\times BR(W^+\rightarrow e^+\nu_e)=144.1\pm 21.2({\rm stat})^{+3.4}_{-10.3}({\rm syst})\pm15\%(norm)$~pb, 
and $\sigma(pp\rightarrow W^{-}X)\times BR(W^-\rightarrow e^-\bar{\nu_e})=31.7\pm 12.1({\rm stat})^{+10.1}_{-8.2}({\rm syst})\pm15\%(norm)$~pb, 
where $BR$ is the branching ratio.
These are shown in Fig.~\ref{fig:xsecfig} and compared to published Tevatron
and S$p\bar{p}$S data.

%%%%%%%%%%%%%%%%%%%%%%%%%%%%%%%%%%%%%%%%%%%%%%%%%%%%%%%%%%%%%%%  Table I
\begin{table}[t]
\caption{Comparison of measured cross sections for electrons and 
positrons with $30<\pt<50$ GeV/$c$ from $W$ and $Z$ decays
with NLO~\cite{deFlorian:2010aa} and NNLO~\cite{Melnikov:2006kv} 
calculations.
The first error is statistical; the second error is systematic 
from the uncertainty in the background; 
the third error is a 15\% normalization uncertainty due to the luminosity 
(10\%), multiple collision (5\%), 
and acceptance and efficiency uncertainties (10\%).
\label{table:xsecresult}}
\begin{ruledtabular}
\begin{tabular}{l c c c}
 & \multicolumn{3}{c}{$\frac{d\sigma}{dy}(30\!<\!p_{T}^e\!<\!50{\rm GeV}/c)|_{y=0}$ [pb]} \\
Lepton & Data & NLO & NNLO\\
\hline
$e^+$ & $50.2 \pm7.2 ^{+1.2}_{-3.6} \pm 15\%$ & 43.2 & 46.8\\
$e^-$ & $9.7 \pm3.7 ^{+2.1}_{-2.5} \pm 15\%$ & 11.3 & 13.5 \\
$e^+$ and $e^-$ & $59.9 \pm8.1 ^{+3.1}_{-6.0} \pm 15\%$ & 54.5 & 60.3\\
\end{tabular}
\end{ruledtabular}
\end{table}

In order to determine the longitudinal spin asymmetry with a
sample of $W$ decays with minimal background contamination, two additional requirements were imposed on the
candidate events.
An isolation cut requiring the sum of cluster energies in the calorimeter
and transverse momenta measured in the drift chamber be less than 2 GeV in
a cone with a radius in $\eta$ and $\phi$ of 0.5 around the candidate track 
was used to remove remaining events with jets.
About 80\% of the signal is kept, while the background is reduced by a factor $\sim\!4$ as shown in 
Fig.~\ref{fig:cy_p}.
The second cut is to reject tracks with $|\alpha|\!<\!1$ mr, which 
reduces charge misidentification to negligible levels.  
There are 42 candidate $W^{+}+Z^{0}$ decays to positrons with a background of $1.7\pm1.0$ 
and 13 candidate $W^{-}+Z^{0}$ decays to electrons with a background of $1.6\pm1.0$ events 
within $30\!<\!\pt\!<\!50$ GeV/$c$ after these two additional cuts. 

The measured asymmetry is given by
\begin{equation}
\epsilon_{L} = \frac{N^{+} - R \cdot N^{-}}{N^{+} + R \cdot N^{-}}
\end{equation}
where $N^{+}$ is the number of events from a beam of positive helicity and 
$N^{-}$ is the number of events from a beam of negative helicity, 
and $R$ is the ratio of the luminosity for the positive and the negative helicity beams. 
The longitudinal spin asymmetry is then calculated from the measured asymmetry according to
\begin{equation}
A_{L} = \frac{\epsilon_{L} \cdot D}{P}
\end{equation}
where $P$ is the beam polarization and $D$ is a dilution correction to account for the 
remaining background in the signal region.

The luminosity-weighted polarization was $0.38 \pm 0.03$ and $0.40 \pm 0.04$ in the two beams.
In RHIC, both beams are bunched, and the bunch helicity alternates
almost every crossing to reduce systematic effects.
The relative luminosities of the four helicity combinations were measured 
by the beam-beam counters, and were all within 1\% of each other.
To treat the low statistics data properly, a likelihood function created 
from the four spin sorted yields corrected by the relative luminosity was used to 
determine the single-spin asymmetry within its physical range [-1,1].

%%%%%%%%%%%%%%%%%%%%%%%%%%%%%%%%%%%%%%%%%%%%%%%%%%%%%%%%%%%%%%%  Fig_3

The measured asymmetries are shown in Table \ref{table:spinasym} for tracks in the background
($12<\pt<20$ GeV/$c$) and signal ($30<\pt<50$ GeV/$c$) regions.
For tracks in the background region,
$\epsilon_{L}$ was found to be zero within uncertainties. 
Significant asymmetries were observed for tracks in the signal region.
The dilution corrections of  
$1.04\pm0.03$ and $1.14\pm0.10$ for positive and negative charges, respectively, 
were applied.

\begin{figure}[t]
\includegraphics[width=1.0\linewidth]{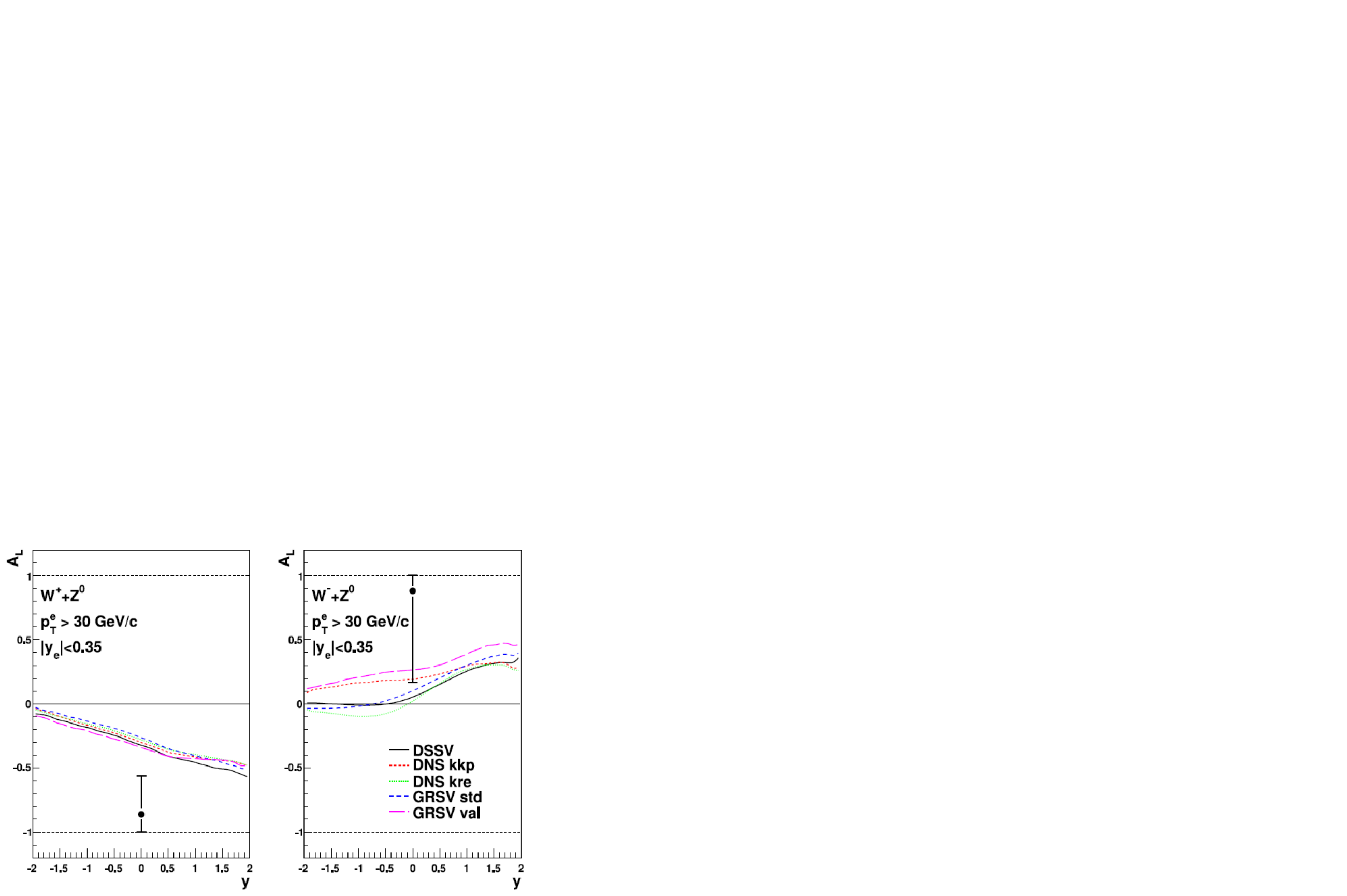}
\caption{(color online) 
 Longitudinal single-spin asymmetries for electrons and positrons 
from $W$ and $Z$ decays. 
 The error bars represent 68\% CL.
The theoretical curves are calculated using NLO with different polarized PDFs~\cite{deFlorian:2010aa}. 
}
\label{fig:alfig}
\end{figure}

Figure~\ref{fig:alfig} compares measured longitudinal single-spin 
asymmetries 
to estimates based on
a sample of polarized PDFs extracted from fits of 
DIS and semi-inclusive DIS data.
The experimental results are consistent with the theoretical 
calculations at 6-15\% confidence level for $A_L^{e^+}$ and 
at 20-37\% for $A_L^{e^-}$.
The observed asymmetries 
are sensitive to the 
polarized quark densities at $x\sim M_W/\sqrt{s} \simeq 0.16$,
and directly demonstrate the parity violating coupling between $W$ bosons 
and light quarks.

%%%%%%%%%%%%%%%%%%%%%%%%%%%%%%%%%%%%%%%%%%%%%%%%%%%%%%%%%%%%%%%  Table II
\begin{table}[t]
\caption{
Longitudinal single-spin asymmetries
\label{table:spinasym}}
\begin{ruledtabular}
\begin{tabular}{l c c c c}
Sample & $\epsilon_{L}$ & $A_L^e(W+Z)$ & 68\%CL & 95\%CL \\
\hline
Bkgrnd + & $-0.015 \pm 0.04$ &  &  &  \\
Signal + & $-0.31 \pm 0.10$ & $-$0.86 & [$-1,-0.56$] & [$-1,-0.16$] \\
\hline
Bkgrnd $-$ & $-0.025 \pm 0.04$ &  &  &  \\
Signal $-$ & $0.29 \pm 0.20$ & +0.88 & [$0.17,1$] & [$-0.60,1$] \\
\end{tabular}
\end{ruledtabular}
\end{table} 

%%%%%%%%%%%%%%%%%%%%%%%%%%%%%%%%%%%%%%%%%%%%%%%%%%%%%%%%%%%%%%%  Fig_4

In summary, we presented first measurements of production cross section 
and nonzero parity violating asymmetry in $W$ and $Z$ production in 
polarized $p+p$ collisions at $\sqrt{s}=500$ GeV.
The results are found to be consistent with theoretical expectations 
and similar measurements of $A_L^{e^\pm}$~\cite{STAR_W}.
RHIC luminosity and PHENIX detector upgrades in progress will make 
it possible in the future to significantly reduce the uncertainties 
for $A_L$ and to extend the measurement to forward rapidity, 
which will improve our knowledge of flavor separated quark and antiquark 
helicity distributions.

% acknowledgments % Run-9 short form for PRL

We thank the Collider-Accelerator Department for 
developing the unique technologies enabling these measurements 
and the Physics Department staff at BNL for vital contributions. 
We also thank D. de Florian, B. Surrow, and J. Balewski for helpful discussions.
We acknowledge support from 
the Office of Nuclear Physics in DOE Office of Science and NSF (U.S.A.), 
MEXT and JSPS (Japan), 
CNPq and FAPESP (Brazil), 
NSFC (China), 
MSMT (Czech Republic),
IN2P3/CNRS and CEA (France), 
BMBF, DAAD, and AvH (Germany), 
OTKA (Hungary), 
DAE and DST (India), 
ISF (Israel), 
NRF (Korea), 
MES, RAS, and FAAE (Russia),
VR and KAW (Sweden), 
U.S. CRDF for the FSU, 
Hungary-US HAESF,
and US-Israel BSF.

%\bibliography{ppg120x0}   

\begin{thebibliography}{26}
\expandafter\ifx\csname natexlab\endcsname\relax\def\natexlab#1{#1}\fi
\expandafter\ifx\csname bibnamefont\endcsname\relax
  \def\bibnamefont#1{#1}\fi
\expandafter\ifx\csname bibfnamefont\endcsname\relax
  \def\bibfnamefont#1{#1}\fi
\expandafter\ifx\csname citenamefont\endcsname\relax
  \def\citenamefont#1{#1}\fi
\expandafter\ifx\csname url\endcsname\relax
  \def\url#1{\texttt{#1}}\fi
\expandafter\ifx\csname urlprefix\endcsname\relax\def\urlprefix{URL }\fi
\providecommand{\bibinfo}[2]{#2}
\providecommand{\eprint}[2][]{\url{#2}}

\bibitem[{\citenamefont{Jaffe and Manohar}(1990)}]{Jaffe:1989jz}
\bibinfo{author}{\bibfnamefont{R.~L.} \bibnamefont{Jaffe}} \bibnamefont{and}
  \bibinfo{author}{\bibfnamefont{A.}~\bibnamefont{Manohar}},
  \bibinfo{journal}{Nucl. Phys.} \textbf{\bibinfo{volume}{B337}},
  \bibinfo{pages}{509} (\bibinfo{year}{1990}).

\bibitem[{\citenamefont{Leader and Anselmino}(1988)}]{Leader:1988vd}
\bibinfo{author}{\bibfnamefont{E.}~\bibnamefont{Leader}} \bibnamefont{and}
  \bibinfo{author}{\bibfnamefont{M.}~\bibnamefont{Anselmino}},
  \bibinfo{journal}{Z. Phys.} \textbf{\bibinfo{volume}{C41}},
  \bibinfo{pages}{239} (\bibinfo{year}{1988}).

\bibitem[{\citenamefont{Kuhn et~al.}(2009)\citenamefont{Kuhn, Chen, and
  Leader}}]{Kuhn:2008sy}
\bibinfo{author}{\bibfnamefont{S.~E.} \bibnamefont{Kuhn}},
  \bibinfo{author}{\bibfnamefont{J.~P.} \bibnamefont{Chen}}, \bibnamefont{and}
  \bibinfo{author}{\bibfnamefont{E.}~\bibnamefont{Leader}},
  \bibinfo{journal}{Prog. Part. Nucl. Phys.} \textbf{\bibinfo{volume}{63}},
  \bibinfo{pages}{1} (\bibinfo{year}{2009}) \bibinfo{note}{and references
  therein.}

\bibitem[{\citenamefont{de~Florian et~al.}(2009)\citenamefont{de~Florian,
  Sassot, Stratmann, and Vogelsang}}]{deFlorian:2009vb}
\bibinfo{author}{\bibfnamefont{D.}~\bibnamefont{de~Florian}},
  \bibinfo{author}{\bibfnamefont{R.}~\bibnamefont{Sassot}},
  \bibinfo{author}{\bibfnamefont{M.}~\bibnamefont{Stratmann}},
  \bibnamefont{and}
  \bibinfo{author}{\bibfnamefont{W.}~\bibnamefont{Vogelsang}},
  \bibinfo{journal}{Phys. Rev.} \textbf{\bibinfo{volume}{D80}},
  \bibinfo{pages}{034030} (\bibinfo{year}{2009}).

\bibitem[{\citenamefont{Alekseev et~al.}(2010)}]{Alekseev:2010}
\bibinfo{author}{\bibfnamefont{M.~G.} \bibnamefont{Alekseev}}
  \bibnamefont{et~al.} (\bibinfo{year}{2010}), \eprint{1007.4061}.

\bibitem[{\citenamefont{Airapetian et~al.}(2005)}]{Airapetian:2004zf}
\bibinfo{author}{\bibfnamefont{A.}~\bibnamefont{Airapetian}}
  \bibnamefont{et~al.}, \bibinfo{journal}{Phys. Rev.}
  \textbf{\bibinfo{volume}{D71}}, \bibinfo{pages}{012003}
  (\bibinfo{year}{2005}).

\bibitem[{\citenamefont{Adeva et~al.}(1998)}]{Adeva:1997qz}
\bibinfo{author}{\bibfnamefont{B.}~\bibnamefont{Adeva}} \bibnamefont{et~al.},
  \bibinfo{journal}{Phys. Lett.} \textbf{\bibinfo{volume}{B420}},
  \bibinfo{pages}{180} (\bibinfo{year}{1998}).

\bibitem[{\citenamefont{Adare et~al.}(2009{\natexlab{a}})}]{Adare:2008px}
\bibinfo{author}{\bibfnamefont{A.}~\bibnamefont{Adare}} \bibnamefont{et~al.},
  \bibinfo{journal}{Phys. Rev. Lett.} \textbf{\bibinfo{volume}{103}},
  \bibinfo{pages}{012003} (\bibinfo{year}{2009}{\natexlab{a}}).

\bibitem[{\citenamefont{Abelev et~al.}(2008)}]{Abelev:2007vt}
\bibinfo{author}{\bibfnamefont{B.~I.} \bibnamefont{Abelev}}
  \bibnamefont{et~al.}, \bibinfo{journal}{Phys. Rev. Lett.}
  \textbf{\bibinfo{volume}{100}}, \bibinfo{pages}{232003}
  (\bibinfo{year}{2008}).

\bibitem[{\citenamefont{Bunce et~al.}(2000)\citenamefont{Bunce, Saito, Soffer,
  and Vogelsang}}]{Bunce:2000uv}
\bibinfo{author}{\bibfnamefont{G.}~\bibnamefont{Bunce}},
  \bibinfo{author}{\bibfnamefont{N.}~\bibnamefont{Saito}},
  \bibinfo{author}{\bibfnamefont{J.}~\bibnamefont{Soffer}}, \bibnamefont{and}
  \bibinfo{author}{\bibfnamefont{W.}~\bibnamefont{Vogelsang}},
  \bibinfo{journal}{Ann. Rev. Nucl. Part. Sci.} \textbf{\bibinfo{volume}{50}},
  \bibinfo{pages}{525} (\bibinfo{year}{2000}).

\bibitem[{\citenamefont{Bourrely and Soffer}(1993)}]{Bourrely:1993dd}
\bibinfo{author}{\bibfnamefont{C.}~\bibnamefont{Bourrely}} \bibnamefont{and}
  \bibinfo{author}{\bibfnamefont{J.}~\bibnamefont{Soffer}},
  \bibinfo{journal}{Phys. Lett.} \textbf{\bibinfo{volume}{B314}},
  \bibinfo{pages}{132} (\bibinfo{year}{1993}).

\bibitem[{\citenamefont{de~Florian and Vogelsang}(2010)}]{deFlorian:2010aa}
\bibinfo{author}{\bibfnamefont{D.}~\bibnamefont{de~Florian}} \bibnamefont{and}
  \bibinfo{author}{\bibfnamefont{W.}~\bibnamefont{Vogelsang}},
  \bibinfo{journal}{Phys. Rev.} \textbf{\bibinfo{volume}{D81}},
  \bibinfo{pages}{094020} (\bibinfo{year}{2010}).

\bibitem[{\citenamefont{Nadolsky and Yuan}(2003)}]{Nadolsky:2003ga}
\bibinfo{author}{\bibfnamefont{P.~M.} \bibnamefont{Nadolsky}} \bibnamefont{and}
  \bibinfo{author}{\bibfnamefont{C.~P.} \bibnamefont{Yuan}},
  \bibinfo{journal}{Nucl. Phys.} \textbf{\bibinfo{volume}{B666}},
  \bibinfo{pages}{31} (\bibinfo{year}{2003}).

\bibitem[{\citenamefont{Aggarwal et~al.}(2010)}]{STAR_W}
\bibinfo{author}{\bibfnamefont{M.~M.} \bibnamefont{Aggarwal}}
  \bibnamefont{et~al.}, \bibinfo{note}{arXiv:1009.0326 [hep-ex]}.

\bibitem[{\citenamefont{Adcox et~al.}(2003)}]{Adcox:2003zm}
\bibinfo{author}{\bibfnamefont{K.}~\bibnamefont{Adcox}} \bibnamefont{et~al.},
  \bibinfo{journal}{Nucl. Instrum. Meth.} \textbf{\bibinfo{volume}{A499}},
  \bibinfo{pages}{469} (\bibinfo{year}{2003}).

\bibitem[{\citenamefont{Adare et~al.}(2009{\natexlab{b}})}]{Adare:2008qb}
\bibinfo{author}{\bibfnamefont{A.}~\bibnamefont{Adare}} \bibnamefont{et~al.},
  \bibinfo{journal}{Phys. Rev.} \textbf{\bibinfo{volume}{D79}},
  \bibinfo{pages}{012003} (\bibinfo{year}{2009}{\natexlab{b}}).

\bibitem[{\citenamefont{Cacciari et~al.}(2005)\citenamefont{Cacciari, Nason,
  and Vogt}}]{Cacciari:2005rk}
\bibinfo{author}{\bibfnamefont{M.}~\bibnamefont{Cacciari}},
  \bibinfo{author}{\bibfnamefont{P.}~\bibnamefont{Nason}}, \bibnamefont{and}
  \bibinfo{author}{\bibfnamefont{R.}~\bibnamefont{Vogt}},
  \bibinfo{journal}{Phys. Rev. Lett.} \textbf{\bibinfo{volume}{95}},
  \bibinfo{pages}{122001} (\bibinfo{year}{2005}); \bibinfo{note}{M. Cacciari,
  private communication}.

\bibitem[{\citenamefont{Adare et~al.}(2006)}]{Adare:2006hc}
\bibinfo{author}{\bibfnamefont{A.}~\bibnamefont{Adare}} \bibnamefont{et~al.},
  \bibinfo{journal}{Phys. Rev. Lett.} \textbf{\bibinfo{volume}{97}},
  \bibinfo{pages}{252002} (\bibinfo{year}{2006}).

\bibitem[{\citenamefont{Sjostrand et~al.}(2001)}]{Sjostrand:2000wi}
\bibinfo{author}{\bibfnamefont{T.}~\bibnamefont{Sj{\"{o}}strand}}
  \bibnamefont{et~al.}, \bibinfo{journal}{Comput. Phys. Commun.}
  \textbf{\bibinfo{volume}{135}}, \bibinfo{pages}{238} (\bibinfo{year}{2001}).

\bibitem[{\citenamefont{Melnikov and Petriello}(2006)}]{Melnikov:2006kv}
\bibinfo{author}{\bibfnamefont{K.}~\bibnamefont{Melnikov}} \bibnamefont{and}
  \bibinfo{author}{\bibfnamefont{F.}~\bibnamefont{Petriello}},
  \bibinfo{journal}{Phys. Rev.} \textbf{\bibinfo{volume}{D74}},
  \bibinfo{pages}{114017} (\bibinfo{year}{2006}).

\bibitem[{\citenamefont{Martin et~al.}(2003)\citenamefont{Martin, Roberts,
  Stirling, and Thorne}}]{Martin:2002aw}
\bibinfo{author}{\bibfnamefont{A.~D.} \bibnamefont{Martin}},
  \bibinfo{author}{\bibfnamefont{R.~G.} \bibnamefont{Roberts}},
  \bibinfo{author}{\bibfnamefont{W.~J.} \bibnamefont{Stirling}},
  \bibnamefont{and} \bibinfo{author}{\bibfnamefont{R.~S.}
  \bibnamefont{Thorne}}, \bibinfo{journal}{Eur. Phys. J.}
  \textbf{\bibinfo{volume}{C28}}, \bibinfo{pages}{455} (\bibinfo{year}{2003}).

\bibitem[{\citenamefont{Martin et~al.}(2009)\citenamefont{Martin, Stirling,
  Thorne, and Watt}}]{Martin:2009iq}
\bibinfo{author}{\bibfnamefont{A.~D.} \bibnamefont{Martin}},
  \bibinfo{author}{\bibfnamefont{W.~J.} \bibnamefont{Stirling}},
  \bibinfo{author}{\bibfnamefont{R.~S.} \bibnamefont{Thorne}},
  \bibnamefont{and} \bibinfo{author}{\bibfnamefont{G.}~\bibnamefont{Watt}},
  \bibinfo{journal}{Eur. Phys. J.} \textbf{\bibinfo{volume}{C63}},
  \bibinfo{pages}{189} (\bibinfo{year}{2009}).

\bibitem[{\citenamefont{Acosta et~al.}(2005)}]{Acosta:2004uq}
\bibinfo{author}{\bibfnamefont{D.~E.} \bibnamefont{Acosta}}
  \bibnamefont{et~al.}, \bibinfo{journal}{Phys. Rev. Lett.}
  \textbf{\bibinfo{volume}{94}}, \bibinfo{pages}{091803}
  (\bibinfo{year}{2005}).

\bibitem[{\citenamefont{Abbott et~al.}(2000)}]{Abbott:1999tt}
\bibinfo{author}{\bibfnamefont{B.}~\bibnamefont{Abbott}} \bibnamefont{et~al.},
  \bibinfo{journal}{Phys. Rev.} \textbf{\bibinfo{volume}{D61}},
  \bibinfo{pages}{072001} (\bibinfo{year}{2000}).

\bibitem[{\citenamefont{Alitti et~al.}(1990)}]{Alitti:1990gj}
\bibinfo{author}{\bibfnamefont{J.}~\bibnamefont{Alitti}} \bibnamefont{et~al.},
  \bibinfo{journal}{Z. Phys.} \textbf{\bibinfo{volume}{C47}},
  \bibinfo{pages}{11} (\bibinfo{year}{1990}).

\bibitem[{\citenamefont{Albajar et~al.}(1989)}]{Albajar:1988ka}
\bibinfo{author}{\bibfnamefont{C.}~\bibnamefont{Albajar}} \bibnamefont{et~al.},
  \bibinfo{journal}{Z. Phys.} \textbf{\bibinfo{volume}{C44}},
  \bibinfo{pages}{15} (\bibinfo{year}{1989}).

\end{thebibliography}

\end{document}